\documentstyle[12pt]{article}

\begin{document}

\title{Time Asymmetry and Chaos in General Relativity.}
\author{Yves Gaspar\\
DAMTP, Centre for Mathematical Sciences,\\
Cambridge University, Wilberforce Road, \\
Cambridge CB3 0WA, UK\\
\it E-mail: yfjmg@yahoo.co.uk}
\maketitle

\abstract{In this work the late-time evolution of Bianchi type $VIII$ models is discussed. These cosmological models exhibit a chaotic behaviour towards the initial singularity and our investigations show that towards the future, far from the initial singularity, these models have a non-chaotic evolution, even in the case of vacuum and without inflation. These space-time solutions turn out to exhibit a particular time asymmetry. On the other hand, investigations of the late-time behaviour of type $VIII$ models by another author have the result that chaos continues for ever in the far future and that these solutions have a time symmetric behaviour: this result was obtained using the approximation methods of Belinski, Khalatnikov and Lifshitz ($BKL$) and we try to find out a possible reason explaining why the different approaches lead to distinct outcomes. It will be shown that, at a heuristic level, the $BKL$ method gives a valid approximation of the late-time evolution of type $VIII$ models, agreeing with the result of our investigations.}

\section{Introduction.}

The fundamental laws of physics are often said to be {\it time
symmetric}: Newton's laws, Electromagnetism, Quantum Mechanics and Einstein's theory of gravity do not distinguish between past and
future. However the phenomena taking place in our world display an {
\it arrow of time}. It is of great interest to try to understand the
physical origin of this {\it time asymmetry}. In this work we will
focus on Einstein's theory of gravity and study how some solutions of the Einstein Field Equations ($EFE$) can distinguish between past and future. 
We will discuss the particular way in which the nature of the solutions 
at early times close to the initial singularity
differs from the nature of the solution at late times, in the far
future.\\

The solutions that will be discussed are the Bianchi type $VIII$ 
models, which are among the most general homogeneous solutions to the $EFE$.
The type $VIII$ model has been studied mainly at early times, where
the possibility of {\it chaotic} behaviour was investigated. This 
solution was found by $BKL$ \cite{BKL},\cite{BKLI},\cite{BKLII} and Misner C. \cite{M} 
to exhibit chaotic mixmaster behaviour as one approaches the 
initial singularity. As clock time $t \rightarrow 0$ the type $VIII$
model evolves through an infinity of Kasner ( Bianchi type $I$ )
stages, in a chaotic way, see \cite{JDB}. The question arises to see 
whether this chaotic mixmaster behaviour continues for ever in the far
future.

P. Halpern \cite{H} has used the methods of V. Belinski, I. M. Khalatnikov 
and I. Lifshitz ($BKL$) \cite{BKL} to study the late-time behaviour of 
type $VIII$ models and the result is that chaos continues for ever in the far future of these models. Our investigation in \cite{BG}, \cite{G} 
shows that this can not be the case. The chaotic mixmaster solution 
close to the initial singularity evolves into a {\it simpler non-chaotic
solution}. This result is consistent with the recent analysis of
Ringstr\"{o}m H. in \cite{R} and other investigations supporting this
result are \cite{WI}, \cite{Ug}, \cite{WII} and \cite{C}.

It is important to understand why the application of the $BKL$ method in \cite{H} 
yields a completely different result. 
Either there could be some problem with the application to the 
late-time regime, or either there could be something intrinsically wrong with 
the $BKL$ method. $BKL$ applied their method also to the study of the 
general inhomogeneous solution at early times close to the initial 
singularity, so it is of relevance to determine what the nature of the 
problem actually is and to see whether the $BKL$ approximation could be used 
in the late-time regime as well. In order to achieve this we will compare in some detail the analysis of P. Halpern in \cite{H} with the investigation of \cite{BG}, \cite{G}.\\

\section{The time evolution of Bianchi type $VIII$ models.}

The Bianchi type $VIII$ universe is one of the four most general models 
of the spatially homogeneous universes and it is an eternally expanding solution. In recent work
it is shown that type $VIII$ universes become the most general of the
spatially homogeneous models when {\it space is compactified}, see 
\cite{BK}. 
One of the reasons to believe that spatial sections of
cosmological models could be compact is that 
formulations of quantum cosmology require the volume of the spatial
sections to be finite, in order for cosmological wave-functions to
exist. So one could expect that it is most probable 
for a compact homogeneous universe to have type $VIII$ symmetry.\\

In \cite{H} Halpern P. argues that the chaotic behaviour characteristic of the early time evolution of type $VIII$ models continues for ever in
the far future and that in this case there would be
no essential difference between past and future for these
models. However our investigation in \cite{BG}, \cite{G} 
shows that the behaviour in the past is very different from that in the future.\\

Let us review how the Bianchi type $VIII$ model evolves at late times, 
far from 
the initial singularity, according to \cite{BG}, \cite{G}, \cite{R}, \cite{WI}, \cite{Ug}, \cite{WII} and \cite{C}. It turns out to be useful to discuss the 
problem in the {\it Hamiltonian formalism} \cite{M}. In this formulation,   
using the so called metric approach \cite{WE}, where the metric
components are the basic variables of the gravitational field, one can
introduce group-invariant and time-independent frame vectors ${e_{a}}$
such that the line element for Bianchi class A models is given by, using an arbitrary time
variable $t^{\prime}$

$${{ds}^2}=-{N(t^{\prime}){d{t^{\prime}}}^2}+{g_{ab}}{W^a}{W^b}$$

where $W^a$ are time-independent one-forms dual to the frame vectors
$e_{a}$. We can introduce three time-dependent scale factors as 

$${g_{ab}}=diag(a^2,b^2,c^2)$$

which can be rewritten as 

\begin{equation}
\label{VIIIbr}
{{g_{ab}}}=diag({e^{2{\beta }_1}},{e^{2{\beta }_2}},{e^{2{\beta }_3}})
\end{equation}
with 
\[
{\beta }_1={{\beta }^0}-2{{\beta }^{+}} 
\]
\[
{\beta }_2={{\beta }^0}+{{\beta }^{+}}+{\sqrt{3}}{{\beta }^{-}} 
\]
\[
{\beta }_3={{\beta }^0}+{{\beta }^{+}}-{\sqrt{3}}{{\beta }^{-}} 
\]
The evolution of the type $VIII$ model corresponds to the motion of a 
universe point in a triangular shaped potential in minisuperspace 
$({{\beta}^{+}},{{\beta}^{-}})$, possessing an infinite open channel
along the ${{\beta}^{+}}$ axis \cite{WE}. The motion of the universe point would then be analogous to motion of a ball in a triangular billiard, the difference being that the reflection angle before the bounce with a wall is not equal to the angle after the bounce. 
If the universe point moves initially in a straight line ( corresponding to Kasner-like behaviour ) it will bounce off a potential wall and then move along another straight line
( a transition to another Kasner-like behaviour ). The triangular
shape of the potential allows the evolution to be {\it chaotic}. As
the model evolves in the future, the Hamiltonian picture implies that
the triangular potential contracts, thus enabling the universe point
to bounce of the walls in a chaotic sequence, ad infinitum, such that
mixmaster behaviour would continue for ever. 
Our analysis based on a combination of the Hamiltonian 
formalism and the orthonormal frame approach shows that this can not be the case. 
The universe point is forced to leave the triangular region
of the potential and to escape along the infinite open channel along
the ${{\beta}^{+}}$ axis, such that ${{\beta}^{-}} \rightarrow
0$ and ${{\beta}^{+}} \rightarrow +\infty$. 
This late-time evolution will be characterised by an infinity of
{\it non-chaotic oscillations} between the two walls of the channel. 
The open channel becomes increasingly narrow as ${{\beta}^{+}} 
\rightarrow +\infty$ and the universe point will exhibit {\it increasingly rapid
non-chaotic oscillations} about the axisymmetric type $VIII$ solution. 
The line element for vacuum type $VIII$ models will tend in this way to
the Bianchi type $III$ form of flat space-time,

\begin{equation}
\label{BIII}
{ds}^2=-{dt}^2+{t^2}({{dx}^2}+{e^{2x}}{{dy}^2})+{{dz}^2} 
\end{equation}
 
( the Bianchi type $VI_h$ diagonal plane wave metric \cite{WE} 
with $h=-1$ reduces to this line element for a particular choice of 
parameters ) such that the shear parameter ${\Sigma}^2$,

$${{\Sigma}^2}={{{\sigma}^2}\over{3{H^2}}}$$

( with ${{\sigma}^2}$ being the shear scalar and $H$ being the Hubble
parameter ), exhibits {\it increasingly rapid non-chaotic 
oscillations}. For non-vacuum perfect fluid models, with equation of state
$p=({\gamma}-1){\rho}$ and ${\gamma} > 2/3$, the mixmaster behaviour
will evolve as well to a {\it simpler non-chaotic solution}
corresponding to the motion of the universe point along the infinite open channel
of the potential. In this case the line element will tend to the 
Collins type $III$ \cite{WE} solution if $2/3 < {\gamma} < 1$ and to
the above vacuum type $III$ line element if $1 \le {\gamma} \le 2$. In
each of these cases the shear parameter 
${{\Sigma}^2}$ exhibits {\it increasingly rapid non-chaotic 
oscillations}. Another remarkable property of these models is that the Weyl scalar \cite{WI}, \cite{WII} is unbounded towards the far future: this might be of relevance in view of the relation between gravitional entropy and the Weyl curvature as conjectured by R. Penrose \cite{RP}. In this context, a further interesting study of the behaviour of the Weyl scalar for type $VIII$ models and for other homogeneous solutions can be found in \cite{SH}.\\

\section{Is there a problem with the $BKL$ approximation method ? }

Let us study the analysis of
Halpern P. \cite{H} and try to understand why this work leads to 
different results.\\
 
First, in this work it is argued that in 
the Hamiltonian picture, the evolution of two neighbouring points 
in the type $VIII$ potential is such that paths will diverge even 
if the points were close to each other initially. This is a feature of
chaotic systems. Furthermore it
is argued that this Hamiltonian system is {\it time reversible} and
that therefore the Bianchi type $VIII$ model is chaotic both towards
and away from the initial singularity. 
Our investigations show that
the {\it time reversibility} can not be used to deduce the behaviour
far from the singularity. The Bianchi type $VIII$ system on the
contrary exhibits {\it time asymmetry}, corresponding to the fact that 
as one approaches the singularity one has {\it chaotic mixmaster} 
behaviour while the far future is characterised
by a simpler {\it non-chaotic} evolution.\\
 
Second, another argument giving different results in the analysis of \cite{H} is the following. In \cite{H} essentially the same method was used as in the work of $BKL$ \cite{BKL}, giving the result that chaos continues for ever in the far future. So does this implies that there is a problem with the $BKL$ method itself or has the method been applied incorrectly ? Or is the $BKL$ approximation method not applicable in the late-time regime ?

In \cite{H} the field equations 

\begin{equation}
\label{VIIIbd}
2({\ln {a}})^{^{\prime \prime }}={(b^2+c^2)^2}-a^4
\end{equation}
\begin{equation}
\label{VIIIbe}
2({\ln {b}})^{^{\prime \prime }}={(a^2+c^2)^2}-b^4
\end{equation}
\begin{equation}
\label{VIIIbf}
2({\ln {c}})^{^{\prime \prime }}={(a^2-b^2)^2}-c^4
\end{equation}
\begin{equation}
\label{VIIIbg}
({\ln {a^2}})^{^{\prime \prime }}+({\ln {b^2}})^{^{\prime \prime 
}}+({\ln {
c^2}})^{^{\prime \prime }}={({\ln {a^2}})^{\prime }}{({\ln 
{b^2}})^{\prime }}
+{({\ln {b^2}})^{\prime }}{({\ln {c^2}})^{\prime }}+{({\ln 
{c^2}})^{\prime }}
{({\ln {a^2}})^{\prime }}
\end{equation}

(the prime denotes differentiation w.r.t. $\eta$ defined by 
$dt=(abc)d{\eta}$) were studied with the
assumption that one of the scale factors is smaller then the other
two, in order to study the evolution of a Kasner stage. 
The transition from one Kasner epoch to the other was parametrised by
writing Kasner exponents as functions of a single parameter $u$ 

$${p_1}={{u}\over{1-u+u^2}}$$

$${p_2}={{1-u}\over{1-u+u^2}}$$

$${p_3}={{u^2-u}\over{1-u+u^2}}$$

with $u > 1$ and such that the scale factors correspond to 

$${a} \sim t^{p_1}$$

$${b} \sim t^{p_2}$$

$${c} \sim t^{p_3}$$

Supposing that initially ${p_1} > {p_3} > 0$ and ${p_{2}} < 0$, and if
one includes only the dominant terms in (\ref{VIIIbd})-(\ref{VIIIbg}), 
we get  
that, as $t \rightarrow +\infty$, the field
equations lead to another Kasner stage with ${{p'}_{3}} > {{p'}_{2}} >
0$ and ${{p'}_{1}} < 0$ and so as the model evolves away from the
singularity, the transition between Kasner epochs is given by the 
transformation

\begin{equation}
\label{VIIIcz}
u \rightarrow u+1
\end{equation}

Two of the scale factors will have exchanged their
increasing and decreasing behaviour while the other scale factor
continues to increase monotonically. 
Now, (\ref{VIIIcz}) means that the parameter
$u$ increases indefinitely, so it will never reach values $ u < 1 $,
which implies that there is {\it no reason for a new Kasner era to
begin} and the alternation between Kasner states seems to continue for
ever. However, as the parameter $u$ continues to increase, the Kasner 
parameters $p_1$, $p_2$ and $p_3$ approach the following values

$${p_{1}} \sim {{1}\over{u}}$$

$${p_{2}} \sim {{-1}\over{u}}$$

$${p_{3}} \sim 1-{{1}\over{u^2}}$$

and thus the exponents $p_1$ and $p_{2}$ become close to each other.
This means that the approximation done at the beginning of this
calculation, namely keeping the dominant scale factor $a$ only in the
right-hand side of (\ref{VIIIbd})-(\ref{VIIIbf}), is not valid any more 
since $a \sim b$.
In \cite{H} the following notation is used to write the resulting field
equations : ${\alpha}={\ln{a}}$, ${\beta}={\ln{b}}$ and 
${\gamma}={\ln{c}}$.
When taking both $a$ and $b$ into account, 
(\ref{VIIIbd})-(\ref{VIIIbf}) can be written as 

$${{\alpha}}^{^{\prime \prime }}+{{\beta}}^{^{\prime \prime }}=0$$

$${{\alpha}}^{^{\prime \prime }}-{{\beta}}^{^{\prime 
\prime}}={e^{4{\beta}}}-{e^{4{\alpha}}}$$

The equations (\ref{VIIIbd})-(\ref{VIIIbg}) also yield a first integral 
which is given by,
when terms containing $c$ are neglected compared to those containing 
dominant contributions of $a$ and $b$,

$${{{\alpha}}^{\prime}}{{{\beta}}^{\prime}}+{{{\beta}}^{\prime}}{{{\gamma}}^{\prime}}+{{{\gamma}}^{\prime}}{{{\alpha}}^{\prime}}={{1}\over{4}}{({e^{2{\alpha}}}+{e^{2{\beta}}})}$$

The subsequent analysis in \cite{H} follows the one of \cite{BKL}: the following change of time variable is made 

\begin{equation}
\label{VIIIda}
{\xi}={{\xi}_{0}}{\exp{[{{2{a_{0}}^2}\over{{{\xi}_{0}}^2}}({\eta}-{{\eta}_{0}})]}}
\end{equation} 

with ${{\xi}_{0}}$, ${a_{0}}$ and ${\eta}_{0}$ being constants.
If we define further ${\chi}={\alpha}-{\beta}$, then the above field 
equations can be written as \cite{H}, \cite{BKL}

\begin{equation}
\label{VIIIdb}
{{\chi}_{{\xi}{\xi}}}+{{1}\over{\xi}}{{\chi}_{{\xi}}}+{{1}\over{2}}{\sinh{2{\chi}}}=0
\end{equation}
 
\begin{equation}
\label{VIIIdc}
{{\gamma}_{\xi}}=-{{1}\over{4{\xi}}}+{{1}\over{8}}{\xi}[{{{\chi}^2}_{\xi}}+{\cosh{2{\chi}}}+1]
\end{equation}
Now it was argued that as ${\eta} \rightarrow +\infty$, we have 
${\xi} \rightarrow +\infty$ because of equation (\ref{VIIIda}), so that 
terms containing ${{1}\over{\xi}}$ can be neglected in (\ref{VIIIdb}) and 
(\ref{VIIIdc} ). If we
assume that $a$ and $b$ are very close to each other, then we have
${\sinh{(2{\chi})}} \approx 2{\chi}$ and these approximations lead to 
the following solution of (\ref{VIIIdb})-(\ref{VIIIdc}) 

\begin{equation}
\label{VIIIdd}
a,b={a_0}{({{\xi}\over{{\xi}_0}})^{{1}\over{2}}}{\exp{[{\pm}({{A}\over{\sqrt{\xi}}}){\sin{({\xi}-{{\xi}_0})}}]}}
\end{equation} 

\begin{equation}
\label{VIIIdda}
c={c_0}{\exp{[{{1}\over{8}}({{\xi}^2}-{{{{\xi}_0}}^2})]}}
\end{equation}

At some stage the increasing scale factor $c$ will become comparable to
$a$ and $b$ and will become the dominant term as 
${\xi} \rightarrow +\infty$ : this enables a {\it new Kasner era} to begin,
with this time $c \gg a,b$. Further developments in \cite{H} are done
in order to show that this type of evolution is chaotic in the far 
future.

Now, the solution (\ref{VIIIdd})-(\ref{VIIIdda}) is valid if ${\xi} 
\rightarrow +\infty$,
which corresponds to ${\eta} \rightarrow +\infty$ because of equation
(\ref{VIIIda}). However as was explained in \cite{BG}, a consequence of 
the inequality $(abc)^{..}>0$ ( which can be deduced directly from the 
field equations (\ref{VIIIbd})-(\ref{VIIIbf}), the dot denotes diferentiation w.r.t. synchronous time $t$ ) can be that

\begin{equation}
\label{BT}
{\eta} \rightarrow 0
\end{equation}

as $t \rightarrow +\infty$. 

We can check this with the future asymptotic
form of the line-element for vacuum Bianchi type $VIII$ that we
obtained, namely the Bianchi type $III$ form of flat space-time given
by (\ref{BIII}). For this line element we have 

$$abc \sim t^2 $$

This means that the relation between the ${\eta}$-time and the
synchronous time $t$ is given by 

$$ {\eta} = {\int{{{dt}\over{abc}}}} \sim {{-1}\over{t}}$$

so that indeed ${\eta} \rightarrow 0$ as $t \rightarrow +\infty$.

This behaviour of $BKL$ ($\eta$) time given by equation (\ref{BT}) has 
also been
observed for type Bianchi $VII_{0}$ models in the work of Wainwright et
al. \cite{WHU}.

This differs however with the derivation of $BKL$ \cite{BKL}, ( see their 
equation 4.18 ) 
where the integration of $dt=(abc)d{\eta}$ yielded 

$$ t \sim {e^{{{1}\over{8}}({{\xi}^2}-{{{{\xi}_0}}^2})}}$$

where ${\xi}_0$ is a constant, because this implies that ${\xi} 
\rightarrow +\infty$
as $t \rightarrow +\infty$. Such behaviour of ${\xi}$ is also supposed 
to be true in the work of I. M. Khalatnikov and V. L. Pokrovsky 
\cite{KP}. This hypothesis is the source of the problem. 
Our result implies that ${\xi} \rightarrow $ constant as 
$t \rightarrow +\infty$, which means that far from the initial
singularity the solution (\ref{VIIIdd})-(\ref{VIIIdda}) is no longer 
valid, since
${{1}\over{\xi}}$-terms are not ignorable in equations
(\ref{VIIIdb})-(\ref{VIIIdc}). Indeed equation (\ref{VIIIdb}) with 
${\sinh{(2{\chi})}} \approx 2{\chi}$ is a Bessel equation, a solution being 
a Bessel function of order zero 

$${\chi}(\xi)=J_{0}(\xi)$$

The fact that ${\xi}$ is assumed to continually increase
is important in the analysis of \cite{H} since this allows the scale 
factor $c$
given by (\ref{VIIIdda}) to become larger then $a$ and $b$, such that 
the
intermediate axisymmetric stage comes to an end in order for a new
Kasner era to develop. As a result it was then explained in \cite{H}
that an "axisymmetric period" with two scale factors close to each other 
is finite in duration and that this ultimately has no effect on the 
solution.   

However if ${\xi} \rightarrow $ constant as
$t \rightarrow +\infty$, then ${\chi}(\xi)=J_{0}(\xi) \rightarrow $ 
constant : the difference between $\alpha$ and $\beta$ will not change. In 
addition, the solution of equation (\ref{VIIIdc}) for ${\gamma}({\xi})$ 
implies that ${\gamma}({\xi}) \rightarrow $ constant as ${\xi} 
\rightarrow $ constant, such that the scale factor $c$ can not become greater 
then $a$ and $b$ as was supposed in \cite{H}.
In other words the axisymmetric regime could never come to an end, which agrees with the result of our investigation and with that of other authors.

Thus summarising one can say that provided one takes into account the 
correct interval for the $BKL$ time variable ${\eta}$, i.e. $]-\infty,0]$, 
then the $BKL$ approximation method predicts that the vacuum Bianchi type 
$VIII$ solution tends to an axisymmetric solution as 
$t \rightarrow +\infty$. Thus as heuristic or approximative analysis 
the $BKL$ method does not fail in the late-time regime.\\

The rigorous analysis of Ringstr\"{o}m H. in \cite{RII} of the Bianchi 
type $IX$ model close to the singularity gave support to the validity of 
the $BKL$ approximation method. 
In fact we have a new support to the heuristic validity of the $BKL$ 
approximation method at late times, since it leads basically to a similar 
result as other rigorous methods \cite{R}, \cite{WI} and \cite{WII}.

\section{Chaos disappears in the far future.}

The fact that chaotic behaviour disappears at late times such that the
type $VIII$ models evolve into a simpler solution even when ${\gamma}
> 2/3$, thus {\it without inflation}, is remarkable in the sense that
for most complex physical systems chaos is a feature of {\it the
whole evolution}, for instance in the case of the three-body problem
there seems to be no time direction in which chaos ceases to
exist. 
In \cite{Lb} it is explained that there
is a frequently overlooked distinction between irreversible and
chaotic behaviour of Hamiltonian systems. The latter does not in 
general appears to possess a
direction of time, i.e. there is no essential distinction between past
and future. By looking at a sequence of "snapshots" or
configurations of the system at different instants of time, one would 
not observe some asymmetry.
The general solutions to the Einstein Field Equations that we discussed do not 
exhibit this distinction between chaos and time asymmetry, since the time 
evolution of the type $VIII$ models distinguishes between past and 
future. The answer to the question "Is the type $VIII$ system chaotic?" should then be: 
"It depends on the direction of time: towards the initial singularity the evolution is chaotic, while in the far future it is non-chaotic". 
The type $VIII$ system appears to loose memory of the chaotic initial state, and evolves in the far future to the axisymmetric type $VIII$ solution. Thus the simple late-time
asymptotic solution can be seen as arising from a wider set of ( chaotic ) initial conditions. 
This time asymmetry corresponds to the one 
defined by Halliwell J. in \cite{TA}, which is based on the distinction 
between the metric at early and late times. 
An important point to note is that the singularity of the type $VIII$ solutions plays a crucial role in the dynamics of these models : some terms in the field equations become 
always negligeable (such that Kasner stages are always possible) close 
to the singularity, while far from the singularity they are not (see 
previous section). This is related to the time asymmetry exhibited by the 
type $VIII$ model. \\

Finally, let us point out that the asymmetry between past and future of type $VIII$ models might be of relevance in the study of the late-time behaviour of cosmological solutions of (super)string theory as studied in \cite{TD}, where mixmaster behaviour was found to occur close to the initial singularity.\\

\section{Conclusion.}

In the work of $BKL$ \cite{BKL} and Halpern \cite{H} the interval of 
variation of the $BKL$ time variable $\eta$ is supposed to be $]-\infty, 
+\infty[$. As a consequence the $BKL$ approximation method predicts mixmaster 
behaviour to occur in vacuum type $VIII$ models 
at late times, far from the initial singularity. In this case the whole 
evolution of the type $VIII$ model would be chaotic. 
If one takes into account that the correct interval of variation for 
$\eta$ is $]-\infty,0[$ then the $BKL$ method predicts that at late times the type 
$VIII$ model tends to the axisymmetric solution, thus in agreement at a 
heuristic level with other rigorous methods. The $BKL$ method is thus a 
valid approximation at late times and it could be applied in order to 
obtain information about the nature of the general vacuum inhomogeneous 
solution in the late-time regime. 
The late-time evolution of type $VIII$ models is non-chaotic, thus they 
distinguish between past and future. Remarkably the type $VIII$ model 
looses memory of the chaotic initial state
and evolves to a simpler non-chaotic future state. One could expect or 
conjecture that a similar time asymmetry will be found in the evolution 
of the general vacuum inhomogeneous solution.
This remarkable type of dynamical feature of the Einstein Field 
Equations is not present in Newtonian or non-relativistic physics and is 
closely related to the occurence of singularities in these space-time 
solutions.

\section{Acknowledgements}
I would like to thank Prof. Arlette Noels from the Institut d'Astrophysique et de Geophysique, University of Liege, Belgium, for giving me the opportunity to present this research at a seminar and I am grateful to Dr. Yves De Rop for the insightfull discussions.

\end{document}